\documentclass[a4paper]{article}

\usepackage{onecolceurws}

\usepackage{amsmath,amssymb,amsthm}
\usepackage{mathpartir}
\usepackage{stmaryrd}
\usepackage{hyperref}
\usepackage{cleveref}
\usepackage[strings]{underscore}
\usepackage{xspace}

\DeclareMathOperator{\Explain}{explain}
\DeclareMathOperator{\Infeasible}{infeasible}
\DeclareMathOperator{\Resolve}{resolve}
\DeclareMathOperator{\Value}{value}
\newcommand{\Basis}{\mathit{Basis}}
\newcommand{\Clauses}{\mathcal{C}}
\newcommand{\rulename}[1]{\texttt{#1}}
\newcommand{\ResT}{Res$^*$\!(T)\xspace}

\newcommand{\ProofRule}[4]{
	\textrm{\rulename{#1:}} & \inferrule{#2}{#3} && \textrm{\textbf{if}} \left\{ \begin{array}{l} #4 \end{array}\right.
}

\newtheorem{theorem}{Theorem}

\title{On the proof complexity of MCSAT}

\author{
	Gereon Kremer \\ RWTH Aachen University \\ gereon.kremer@cs.rwth-aachen.de
\and
	Erika Ábrahám \\ RWTH Aachen University \\ eab@cs.rwth-aachen.de
\and
	Vijay Ganesh \\ University of Waterloo \\ vijay.ganesh@uwaterloo.ca
}
\institution{}
	
\begin{document}
\maketitle
	
\begin{abstract}
	Satisfiability Modulo Theories (SMT) and SAT solvers are critical
	components in many formal software tools,
	primarily due to the fact that they are able to easily solve
	logical problem instances with millions of variables and clauses. This
	efficiency of solvers is in surprising contrast to the traditional
	complexity theory position that the problems that these solvers address
	are believed to be hard in the worst case. In an attempt to resolve
	this apparent discrepancy between theory and practice, theorists
	have proposed the study of these solvers as proof systems that would
	enable establishing appropriate lower and upper bounds on their
	complexity. For example, in recent years it has been shown that (idealized models of) SAT
	solvers are polynomially equivalent to the general resolution proof
	system for propositional logic, and SMT solvers that use the CDCL(T) architecture are
	polynomially equivalent to the \ResT proof system.

	In this paper, we extend this program to the MCSAT approach for SMT
	solving by showing that the MCSAT architecture is polynomially
	equivalent to the \ResT proof system. Thus, we establish an equivalence between
	CDCL(T) and MCSAT from a proof-complexity theoretic point of
	view. This is a first and essential step towards a richer theory
	that may help (parametrically) characterize the kinds of formulas for
	which MCSAT-based SMT solvers can perform well.

\end{abstract}
\vskip 32pt

\section{Introduction}

In this work we are interested in proof systems to decide the satisfiability of (quantifier-free first-order logic) formulas.
Well-known proof systems include variants of the \emph{resolution proof system} for propositional logic, but also for first-order logic as presented in \cite{Robere2018}.
An interesting measure to compare proof systems is their \emph{proof complexity}: if we can show that in a proof system we can generate shorter proofs (with less proof steps) than in another, we consider the former proof system \emph{more powerful}.

One particular problem that concerns quantifier-free first-order logic formulas is to determine their satisfiability.
Answering this question (at least for certain theories) has spawned an active field of research called \emph{satisfiability modulo theories (SMT) solving} \cite{handbook,decision_proc}.
The predominant approach is called CDCL(T) and works by combining a solver for propositional logic (a \emph{SAT solver}) with a solver that checks a set of theory constraints for consistency (a \emph{theory solver}).

A significantly different solving technique called MCSAT was presented in \cite{Jovanovic2012,deMoura2013} which provides a significantly tighter integration of the Boolean and the theory reasoning.
This solver performs extremely well in practice, at least for ``hard'' theories like non-linear real arithmetic, thus naturally raising the question whether this MCSAT implementation only happens to include better heuristics, or whether MCSAT itself has some more fundamental advantage over CDCL(T).

We tackle this question from the perspective of proof complexity, allowing us to abstract from certain practical aspects very naturally, for example the impact of heuristics.
We can understand both CDCL(T) and MCSAT as proof systems \cite{Nieuwenhuis2006,deMoura2013}. The proof complexity of CDCL(T) was found in \cite{Robere2018} to be equivalent to the \ResT proof system.
In this paper, we show that also the MCSAT architecture is polynomially equivalent to the \ResT proof system. Thus, we establish an equivalence between CDCL(T) and MCSAT from a complexity-theoretic
view.

It is important to realize, that there is a subtle but important difference between what we call DPLL -- referring to the original DPLL method from \cite{Davis1962} -- and the version extended with clause learning -- proposed in \cite{Silva1996} -- that we now call CDCL.
Compared to CDCL(T) (and \ResT), DPLL(T) is weaker in terms of proof complexity and we only argue about CDCL(T) throughout this paper.

\section{Preliminaries}

\subsection{Proof Systems}

We assume the reader to be familiar with deductive proof systems and give in the following just a short introduction to notation. For readability, in the following we will use \emph{conditional} proof rules of the form
\vspace*{-1ex}
\begin{align*}
	\ProofRule{PR}{A_1\ \ldots\ A_n}{C}{\texttt{S}_1, \ldots,\ \texttt{S}_m}
\end{align*}
\noindent to form proof systems. Such a rule with name \rulename{PR} allows to derive the consequent $C$ from the antecedents $A_1,\ldots,A_n$ \emph{and some side conditions} $\texttt{S}_1,\ldots,\texttt{S}_m$. We will typically describe some \emph{state} in a single antecedent $A$ and some conditions on this state in $\texttt{S}_1,\ldots,\texttt{S}_m$. Note that the above conditional proof rule is equivalent to one without side-condition but with the antecedents $A_1,\ldots,A_n, \texttt{S}_1,\ldots,\texttt{S}_m$.

A \emph{(deductive) proof system} is a set of such (conditional) proof rules. By a \emph{derivability statement} $\Gamma \vdash  C$ we denote that we can \emph{prove} $C$ from a set $\Gamma$ of antecedents (and side conditions), i.e., that $C$ is derivable from $\Gamma$ by finitely many proof rule applications. By $\Gamma\models C$ we denote that $\Gamma$ assures the truth of $C$ (with respect to an underlying semantics). Qualitative properties of proof systems include soundness (if $\Gamma\vdash C$ then $\Gamma\models C$) and completeness (if $\Gamma\models C$ then $\Gamma\vdash C$). All the proof systems that we consider in this paper are sound and complete; we refer to the respective references for details.


The \emph{length} of a proof in a proof system is the number of proof rule applications used in it; a proof is \emph{shorter} then another one if its length is smaller. The \emph{proof complexity}  for $\Gamma \vdash  C$ is the length of the \emph{shortest} proof that derives $C$ from $\Gamma$.
Our aim in this paper is to compare for certain classes of derivability statements their proof complexities in different proof systems. 

Assume two proof systems $P_1$ and $P_2$, and a metric to measure the size of derivability statements that are true in both systems. We say that $P_1$ is \emph{more powerful} than $P_2$ on a set of such statements if the proof complexity grows (at most) polynomially with the statement (input) size in $P_1$ but exponentially in $P_2$.
We call $P_1$ and $P_2$ \emph{comparable} if one of them is more powerful than the other on \emph{all} such classes, and \emph{incomparable} otherwise.
We refer to \cite{Robere2018} for a discussion of existing proof systems and their relations in terms of proof complexity.

\subsection{Model-constructing Satisfiability Calculus (MCSAT)}

In the following we recall the MCSAT proof system for the
satisfiability check of quantifier-free first-order logic formulas as
defined in \cite{deMoura2013}.  Following \cite{deMoura2013}, we present the proof rules in three categories: \emph{Boolean reasoning} (\cref{fig:mcsatA}), \emph{conflict analysis} (\cref{fig:mcsatB}) and \emph{theory reasoning} (\cref{fig:mcsatC}).
The \emph{Boolean reasoning} and \emph{conflict analysis} parts essentially constitute a regular CDCL-style SAT solver while the \emph{theory reasoning} part enhances the proof system to allow for SMT-style theory computations.

The only modification to \cite{deMoura2013} is the addition of the
\rulename{Restart} rule (\cref{fig:mcsatA}), which is sound and also does not introduce non-termination as long as we ensure that it is applied with increasing
periodicity \cite{Nieuwenhuis2006}.  Restarts are a standard technique
both in CDCL-style SAT solving and in SMT solving, but such a rule was
not included in the original MCSAT proof system.  However, we will
need this rule for the proof complexity results in the next section.

Due to space restrictions, in the following we discuss the rules only briefly and
refer to \cite{deMoura2013} for more detailed explanations.

\begin{figure}[t]
	\begin{align*}
		\ProofRule{Decide}{\langle M, \Clauses \rangle}{\langle \llbracket M, L \rrbracket, \Clauses \rangle}{
			L \in \Basis \\
			\Value(L,M) = \textit{undef}
		} \\[1ex]
		\ProofRule{Propagate}{\langle M, \Clauses \rangle}{\langle \llbracket M,C \rightarrow L \rrbracket, \Clauses \rangle}{
			C = (L_1 \lor \dots \lor L_n \lor L) \in \Clauses, \\
			\forall i. \Value(L_i,M) = \textit{false}, \Value(L,M) = \textit{undef}
		} \\[1ex]
		\ProofRule{Conflict}{\langle M, \Clauses \rangle}{\langle M, \Clauses \rangle \Vdash C}{
			C = (L_1 \lor \dots \lor L_n)\in \Clauses,\\
			\forall i.\Value(L_i,M) = \textit{false}
		} \\[1ex]
		\ProofRule{Sat}{\langle M, \Clauses \rangle}{\textrm{\texttt{SAT}}}{
			M \textit{ is complete},\\
			\forall C=(L_1\vee\ldots\vee L_n)\in\Clauses.\exists i. \Value(L_i,M) = \textit{true}
		} \\[1ex]
		\ProofRule{Forget}{\langle M, \Clauses \rangle}{\langle M, \Clauses \setminus \{ C \} \rangle}{
			C \in \Clauses \textit{ is a learned clause}
		} \\[1ex]
		\ProofRule{Restart}{\langle M, \Clauses \rangle \Vdash C}{\langle \llbracket \rrbracket, \Clauses \rangle}{true}
	\end{align*}

	\caption{The MCSAT proof rules I: Boolean reasoning}
	\label{fig:mcsatA}
\end{figure}

\begin{figure}
	\begin{align*}
		\ProofRule{Resolve}{\langle \llbracket M, D \rightarrow L \rrbracket, \Clauses \rangle \Vdash C}{\langle M, \Clauses \rangle \Vdash R}{
			\neg L \in C, \\
			R = \Resolve(C,D,L)
		} \\[1ex]
		\ProofRule{Consume$_1$}{\langle \llbracket M, D \rightarrow L \rrbracket, \Clauses \rangle \Vdash C}{\langle M, \Clauses \rangle \Vdash C}{
			\neg L \not\in C
		} \\[1ex]
		\ProofRule{Consume$_2$}{\langle \llbracket M, L \rrbracket, \Clauses \rangle \Vdash C}{\langle M, \Clauses \rangle \Vdash C}{
			\neg L \not\in C
		} \\[1ex]
		\ProofRule{Backjump}{\langle \llbracket M, N \rrbracket, \Clauses \rangle \Vdash C}{\langle \llbracket M, C \rightarrow L \rrbracket, \Clauses \rangle}{
			C = (L_1 \lor \dots \lor L_n \lor L), \\
			\forall i. \Value(L_i,M) = \textit{false}, \\
			\Value(L,M) = \textit{undef}, \\
			N \textit{ starts with a decision}
		} \\[1ex]
		\ProofRule{Unsat}{\langle M, \Clauses \rangle \Vdash \textit{false}}{\textrm{\texttt{UNSAT}}}{
			\textit{true}
		} \\[1ex]
		\ProofRule{Learn}{\langle M, \Clauses \rangle \Vdash C}{\langle M, \Clauses \cup \{C\} \rangle \Vdash C}{
			C \not\in \Clauses
		}
	\end{align*}

	\caption{The MCSAT proof rules II: conflict analysis}
	\label{fig:mcsatB}
\end{figure}

The proof system works on states of the form $\langle M, \Clauses \rangle$ consisting of a \emph{trail} $M$ and a set $\Clauses$ of clauses  whose satisfaction we are interested in, where the literals of the clauses are constraints from a given theory $T$. The trail is a list of Boolean decisions, theory decisions, Boolean propagations and theory propagations. Decisions represent exploration by enumeration: a Boolean decision $L$ assigns the Boolean value \textit{true} to the literal $L$ (\cref{fig:mcsatA}, rule \rulename{Decide}; the $\Basis$ set and the $\Value$ function will be explained a bit later), whereas a theory decision $x\mapsto \alpha_x$ assigns the theory value $\alpha_x$ to the theory variable $x$ (\cref{fig:mcsatC}, rule \rulename{T-Decide}; consistency of a trail is explained below). A Boolean propagation $C\rightarrow L$ states that the satisfaction of $\Clauses$ under previous decisions is possible only if $L$ is \textit{true} with the reason being the clause $C$ (\cref{fig:mcsatA}, rule \rulename{Propagate}). Analogously, a theory propagation $E\rightarrow L$ states an implication based on a lemma, i.e. a tautology in the theory (\rulename{T-Propagate}, \cref{fig:mcsatC}). We write $\llbracket M_1, \dots, M_k \rrbracket$ to explicitly denote individual elements of the trail, and we call a trail \emph{complete} if it assigns a Boolean value to each variable or its negation (either by Boolean decision or propagation) as well as a theory value to each theory variable.
Satisfiability of a CNF formula can be proven by deriving the \texttt{SAT} state, based on a complete trail that satisfies each clause (\cref{fig:mcsatA}, rule \rulename{Sat}).

For theory variables $x$, if there is some theory value $\alpha_x$ such that $x\mapsto \alpha_x$ is in the trail $M$ then the value $\Value(x,M)$ of a theory variable $x$ in trail $M$ is $\alpha_x$, and $\textit{undef}$ otherwise. 

A literal $L$, which is either a theory constraint $c$ or the negation $\neg c$ of a constraint $c$, can  be evaluated by two semantics: $c$ or $\neg c$ can enter the trail by Boolean decision or propagation, but also the theory variables in $c$ can be assigned a value from the theory domain. The proof system argues in both the Boolean and the theory domain concurrently, but it assures that they stay \emph{consistent}, meaning that the Boolean and the theory evaluation of a constraint never contradict. Thus the value $\Value(L,M)$ of a literal $L$ in a trail $M$ is $\textit{true}$ if $L$ is decided or propagated in $M$, or if $L$ evaluates to true when we substitute the theory values from $M$ in $L$; it is \textit{false} if $\neg L$ is decided or propagated in $M$, or if $L$ evaluates to \textit{false} under the theory assignments; and \textit{undef} otherwise. We call a consistent trail $M$ \emph{feasible} if the trail's theory variable assignments can be extended to satisfy all clauses $C \in \Clauses$ with $\Value(C,M)=\textit{true}$; by $\Infeasible(M)$ we denote that $M$ is not feasible.

\begin{figure}[t]
	\begin{align*}
		\ProofRule{T-Propagate}{\langle M, \Clauses \rangle}{\langle \llbracket M, E \rightarrow L \rrbracket, \Clauses \rangle}{
			L \in \Basis, \\
			\Value(L,M) = \textit{undef}, \\
			\Infeasible(\llbracket M, \neg L \rrbracket), \\
			E = \Explain(\llbracket M, \neg L \rrbracket) 
		} \\[1ex]
		\ProofRule{T-Decide}{\langle M, \Clauses \rangle}{\langle \llbracket M, x \mapsto \alpha_x \rrbracket, \Clauses \rangle}{
			x \textit{ is a theory variable in } \Clauses,\\
			\Value(x,M) = \textit{undef}, \\
			\llbracket M, x \mapsto \alpha_x \rrbracket \textit{ is consistent}
		} \\[1ex]
		\ProofRule{T-Conflict}{\langle M, \Clauses \rangle}{\langle M, \Clauses \rangle \Vdash E}{
			\Infeasible(M), \\
			E = \Explain(M)
		} \\[1ex]
		\ProofRule{T-Consume}{\langle \llbracket M, x \mapsto \alpha_x \rrbracket, \Clauses \rangle \Vdash C}{\langle M, \Clauses \rangle \Vdash C}{
			\Value(C,M) = \textit{false}
		} \\[1ex]
		\ProofRule{T-Backjump-Decide}{\langle \llbracket M, x \mapsto \alpha_x, N \rrbracket, \Clauses \rangle \Vdash C}{\langle \llbracket M, L \rrbracket, \Clauses \rangle}{
			C = (L_1 \lor \dots \lor L_n \lor L), \\
			\exists i. \Value(L_i,M) = \textit{undef}, \\
			\Value(L,M) = \textit{undef}
		}
	\end{align*}

\caption{The MCSAT proof rules III: theory reasoning}
  \label{fig:mcsatC}
\end{figure}

From a \emph{search state} $\langle M, \Clauses \rangle$ we enter the \emph{conflict state} $\langle M, \Clauses \rangle \Vdash C$ when a violated \emph{conflict clause} (i.e., a clause whose literals all evaluate to \textit{false} under the current trail $M$) is detected (\cref{fig:mcsatA}, rule \rulename{Conflict}), or the conflict state  $\langle M, \Clauses \rangle \Vdash E$ when the trail is infeasible and $E$ is a theory lemma explaining this fact.
The rules \rulename{Resolve}, \rulename{Consume$_1$}, \rulename{Consume$_2$}, \rulename{T-Consume}, \rulename{Backjump}, \rulename{T-Backjump-Decide} and \rulename{Learn} allow to implement conflict resolution in the style of conflict-driven clause learning (CDCL), where $\textit{resolve}(C,D,L)$ denotes the result of the (propositional) resolution applied to the clauses $C$ and $D$ with respect to the literal $L$.
Unsatisfiability of a CNF formula can be proven by deriving the \texttt{UNSAT} state from the empty trail, based on the derivation of the empty (\textit{false}) clause (\cref{fig:mcsatB}, rule \rulename{Unsat}).

The $\Explain$ function explains theory-specific propagations and infeasible states.
The proof systems assumes the existence of a \emph{finite} set $\Basis$ of theory constraints such that all literals from all clauses, especially those returned by the $\Explain$ function, are contained in it. 
The finiteness provides a nice termination argument as only finitely many (different) clauses exist.
Also explanations can be learned and forgotten (rules \rulename{Learn} in \cref{fig:mcsatB} and \rulename{Forget} in \cref{fig:mcsatA}).

\subsection{The Proof Systems Res(T) and \ResT}

Our reference proof system is (one of the variants of) the resolution proof system.
Two resolution proof systems for first-order logic with some theory T called Res(T) and \ResT are discussed in \cite{Robere2018}.
They enhance the traditional resolution proof system for propositional logic by a proof rule that allows for the introduction of new clauses that state tautologies from the theory.
While \ResT allows for \emph{strong theory derivation} which may also introduce new literals (theory constraints), Res(T) is restricted to \emph{(regular) theory derivation} that allows only literals that occur in the formula already.

Introducing new literals can make a significant difference for certain problem classes as discussed in \cite{Robere2018}.
As the MCSAT proof system may as well introduce new literals -- the $\Explain$ method makes heavy use of this in many cases -- we use \ResT for the following comparison, consisting of the \rulename{Resolution} rule and the \rulename{Strong Theory Derivation} rule.

Note that these proof systems lack dedicated final states -- like \texttt{SAT} and \texttt{UNSAT} from MCSAT.
We implicitly derive \texttt{UNSAT} if \rulename{Resolution} produces the empty clause and \texttt{SAT} if no new clauses can be derived (with \rulename{Resolution} or \rulename{(Regular) Theory Derivation}), thus essentially when a fixed point is reached. Note that this proof system does not (directly) allow to extract a satisfying assignment.

\begin{figure}[t]
	\begin{align*}
		\ProofRule{Resolution}{(C \lor l) \quad (D \lor \neg l)}{(C \lor D)}{
			\textit{true}
		} \\[1ex]
		\ProofRule{(Regular) Theory Derivation}{\qquad\varphi\qquad}{\varphi \land C}{
			T \models C, \\
			l \in \varphi \textrm{ for all } l \in C
		} \\[1ex]
		\ProofRule{Strong Theory Derivation}{\qquad\varphi\qquad}{\varphi \land C}{
			T \models C
		}
	\end{align*}

	\caption{The Res(T) and \ResT proof systems}\label{fig:res}
\end{figure}

\section{Content}

We now state and afterwards prove our core theorem.
\begin{theorem}\label{thm:the-theorem}
	The \ResT proof system and the MCSAT proof system are \emph{equivalent} with respect to their proof complexity on \emph{first-order logic with any theory}.
\end{theorem}

We give our proof in two steps: first we show that MCSAT can simulate \ResT{} -- meaning that for any \ResT proof MCSAT can construct a proof that is at most polynomially longer  -- and finally show the reverse statement, that \ResT can simulate MCSAT.
What we show is actually a slightly stronger statement: instead of constructing \emph{some} proof (that is at most polynomially longer) we construct a logically equivalent proof, yielding something we could describe as \emph{algorithmic equivalency}. Of course these proofs will not be \emph{syntactically identical}, but they describe \emph{logically equivalent} proof steps.

\subsection{MCSAT simulates \ResT}

To show that MCSAT can simulate \ResT, we need to show that MCSAT can simulate both the \rulename{Resolution} rule and the \rulename{Strong Theory Derivation} rule with at most polynomial overhead.

\paragraph{\rulename{Resolution}.}

Assuming that our set of clauses $\Clauses$ contains the clauses $(C \lor L)$ and $(D \lor \neg L)$, we need to add $(C \lor D)$ to $\Clauses$.
Let us first handle a special case. Assume that we have a literal in $C$ whose negation is in $D$. In this case $(C \lor D) \equiv \textit{true}$ and there is nothing to do.
Similarly, if $(C \lor D) \in \Clauses$ then we do not need to add it.
From here on we assume that $(C \lor D) \not\equiv \textit{true}$ and $(C \lor D) \not\in \Clauses$.
Starting from an empty trail, the clause $(C \lor D)$ can be learned using the MCSAT proof rules as follows:

\vspace*{-2ex}
\begin{align*}
	\intertext{
		First we apply the \rulename{Decide} rule for all literals $L_1,\ldots,L_n$ of $C$ and $D$. Note that we start with the empty trail, thus initially all literals are undefined and can therefore be decided. Note furthermore that we assumed a finite $\Basis$ set that contains all literals from all clauses in $\Clauses$.
	}
	\ProofRule{Decide}{\langle M, \Clauses \rangle}{\langle \llbracket M, \neg L_i \rrbracket, \Clauses \rangle}{
		L_i \in \Basis,\\ \Value(L_i,M) = \textit{undef}
	}
	\intertext{
		Now, the trail evaluates both $C$ and $D$ to \textit{false} and we use the \rulename{Propagate} rule with $(C \lor L)$ to propagate $L$.
	}
	\ProofRule{Propagate}{\langle M, \Clauses \rangle}{\langle \llbracket M, (C \lor L) \rightarrow L \rrbracket, \Clauses \rangle}{
		\Value(C,M) = \textit{false}, \\
		\Value(L,M) = \textit{undef}
	}
	\intertext{
		Having $\Value(L,M) = \textit{true}$ now, $(D \lor \neg L)$ is conflicting and we apply the \rulename{Conflict} rule.
	}
	\ProofRule{Conflict}{\langle \llbracket M, (C \lor L) \rightarrow L \rrbracket, \Clauses \rangle}{\langle \llbracket M, (C \lor L) \rightarrow L \rrbracket, \Clauses \rangle \Vdash (D \lor \neg L)}{
		(D \lor \neg L) \in \Clauses,\\
		\Value(D \lor \neg L) = \textit{false}
	}
	\intertext{
		We perform resolution using the \rulename{Resolve} rule to obtain $(C \lor D)$.
	}
	\ProofRule{Resolve}{\langle \llbracket M, (C {\lor} L) {\rightarrow} L \rrbracket, \Clauses \rangle \Vdash (D {\lor} \neg L)}{\langle M, \Clauses \rangle \Vdash (C {\lor} D)}{
		L \in (C {\lor} L), \\
		(C{\lor}D) = \Resolve( \\
		\quad (C {\lor} L),(D {\lor} \neg L),L)
	}
	\intertext{
		To add the conflict clause to the set of clauses $\Clauses$ we use the \rulename{Learn} rule.
	}
	\ProofRule{Learn}{\langle M, \Clauses \rangle \Vdash (C \lor D)}{\langle M, \Clauses \cup \{(C \lor D)\} \rangle \Vdash (C \lor D)}{
		(C \lor D) \not\in \Clauses
	}
	\intertext{
		We have achieved our goal of adding $(C \lor D)$ to $\Clauses$ and return to the initial state with the \rulename{Restart} rule.
	}
	\ProofRule{Restart}{\langle M, \Clauses \cup \{(C \lor D)\}  \rangle \Vdash (C \lor D)}{\langle \llbracket \rrbracket, \Clauses \cup \{(C \lor D)\}  \rangle}{
		\textit{true}
	}
\end{align*}

We observe that this sequence of proof rules is polynomial in the size of the clause $(C \lor D)$ -- we need $|C| + |D| + 5$ rule applications -- and we return to the same initial state afterwards, except for the added clause $(C \lor D)$.
Note that we did not use any theory reasoning in the MCSAT rule applications and thus pay no \emph{hidden costs} (as we later discuss in \cref{sec:theory-computations}).

\paragraph{\rulename{Strong Theory Derivation}.}

We need to create some arbitrary clause $C$ which is valid in the theory, that is $T \models C$.
Similar to the previous simulation, we assume that $C \not\equiv \textit{true}$ and $C \not\in \Clauses$ as there is nothing to do in these cases.

Our main hurdle is that MCSAT does not allow for learning arbitrary clauses but only the current conflict clause.
We therefore have to construct an (artificial) conflict that yields the desired clause.
We assume that our finite basis $\Basis$ includes all literals that ever occur in the \ResT proof.
We can construct and learn an arbitrary (valid) clause $C$ using the MCSAT proof rules as follows:

\vspace*{-2ex}
\begin{align*}
	\intertext{
		Starting again from the empty trail, we use \rulename{Decide} repeatedly to add $\neg L$ for every $L \in C$ to the trail.
		Note that \rulename{Decide} allows to decide any literal from $\Basis$, independent on whether they appear in the input formula.
	}
	\ProofRule{Decide}{\langle M, \Clauses \rangle}{\langle \llbracket M, \neg L \rrbracket, \Clauses \rangle}{
		L \in \Basis\\
		\Value(L,M) = \textit{undef}
	}
	\intertext{
		We know that $C$ is a valid clause ($T \models C$) but we have $\Value(C,M) = \textit{false}$ due to the previous decisions.
		Thus $M$ is infeasible and we can apply the \rulename{T-Conflict} rule with $E = C$.
		Recall that a trail is \emph{infeasible} if it is inconsistent in the theory -- which is the case here as $M$ implies $\neg C$ but $T \models C$.
	}
	\ProofRule{T-Conflict}{\langle M, \Clauses \rangle}{\langle M, \Clauses \rangle \Vdash C}{
		\Infeasible(M),\\
		C = \Explain(M)
	}
	\intertext{
		Now we can learn the desired clause $C$ using the \rulename{Learn} rule.
	}
	\ProofRule{Learn}{\langle M, \Clauses \rangle \Vdash C}{\langle M, \Clauses \cup \{C\} \rangle \Vdash C}{
		C \not\in \Clauses
	}
	\intertext{
		Finally we return to the initial state with the \rulename{Restart} rule.
	}
	\ProofRule{Restart}{\langle M, \Clauses \cup \{C\} \rangle \Vdash C}{\langle \llbracket \rrbracket, \Clauses \cup \{C\} \rangle}{
		\textit{true}
	}
\end{align*}
We observe that we need $|C| + 3$ proof rule applications to learn an arbitrary clause (that is neither $\textit{true}$ nor already present in $\Clauses$) and return to the initial state.

\subsubsection{Practicability}

We have seen that the MCSAT framework assumes that it can decide the feasibility of the trail, though (depending on the theory) deciding (in)feasibility might be computationally hard or even undecidable.
While theorists might simply assume an ``oracle'' we now examine how actual implementations deal with this.
Practically relevant implementations of $\Infeasible(M)$ for non-linear real arithmetic are incomplete. The respective papers suggest to apply the \rulename{T-Conflict} rule only if infeasibility can be detected by checking the consistency of \emph{univariate} constraints.
One might even (reasonably) argue that this restriction is not a technical one, but is a fundamental idea in MCSAT that allows the theory exploration in the first place -- a complete implementation of $\Infeasible(M)$ would essentially prevent the theory exploration from happening (like in our simulation).
Simulating the \rulename{Strong Theory Derivation} rule with an incomplete implementation of $\Infeasible(M)$ is a bit technical but doable, however, we need to consider also the length of the resulting proof.

The basic idea of the theory exploration in MCSAT is to guess a partial theory assignment (via \rulename{T-Decide}) until its infeasibility can be detected, generalize the unsatisfying assignment to an unsatisfying region around it (via \rulename{T-Conflict}) and exclude it (via \rulename{Learn}) from the further search and backtrack the theory assignment. This usually happens multiple times for a certain variable until the whole space is excluded for this variable and we have to change the assignment of the previously assigned theory variable.

In the above scenario the trail is already unsatisfiable due to the Boolean assignments, and we eventually discover this fact after exploring a certain number of regions and come up with a reason for the conflict.
Note that the number of rule applications needed in this case grows with the number of regions we need to exclude and the complexity of this process highly depends on the background theory.
For the question of completeness we refer to \cite{deMoura2013} while we observe that for non-linear arithmetic, considering the number of cells a cylindrical algebraic decomposition may have to consider, the number of regions may easily grow exponentially (e.g. in the number of theory variables).

This unfortunately conflicts with our hope to find a \emph{polynomial} reduction. However, this was to be expected: a major aspect of MCSAT is that certain parts of the theory reasoning are moved into the core proof system. Thus we should not be surprised if our core proof system exhibits bad asymptotic behavior if we consider a hard theory, while the proof system we compare it to completely \emph{hides} the theory reasoning from the complexity measure. Note that theory reasoning for non-linear arithmetic based on cylindrical algebraic decomposition, whether it is $\Infeasible(M)$ in MCSAT or the question whether $T \models C$ in \ResT may incur a doubly exponential runtime cost.
We discuss how these two theory questions relate in \cref{sec:theory-computations}.

We thus conclude that the presented reduction is polynomial for the MCSAT proof system, but may not be for an actual MCSAT implementation.
The additional complexity is however not \emph{new} but only \emph{becomes visible} as the MCSAT proof system makes the theory reasoning explicit while \ResT does not.

\subsection{\ResT simulates MCSAT}

We observe that MCSAT has three separate places where clauses can ``exist'', namely the set of clauses $\Clauses$, the trail $M$ and the current conflict clause $C$.
When simulating MCSAT with \ResT, we make sure that the set of clauses that \ResT operates on always includes $\Clauses$, clauses from $M$ and $C$.

Note that \ResT retains all clauses that it constructs as is not designed to \emph{forget} clauses while MCSAT may drop clauses occasionally.
We note that removing clauses can bring advantages in practice, but the number of additional clauses is linear in the number of rule applications -- MCSAT needs at least one rule application to construct a clause in the first place -- and the practical overhead of additional clauses -- for example due to larger lookup tables -- is polynomial.

To prove that \ResT simulates MCSAT, it suffices to show that all clauses that ever occur in the MCSAT derivation can also be constructed using the \ResT proof rules.
In case of unsatisfiability \ResT deduces the empty clause immediately before the application of the \rulename{Unsat} rule in MCSAT while any other termination of the \ResT proof system indicates satisfiability.
We assume that initially both proof systems start with the same set of input clauses.
We identify the rules where the MCSAT rule system constructs new clauses: \rulename{Resolve}, \rulename{T-Propagate} and \rulename{T-Conflict}.

All other rules either do not manipulate any clauses (\rulename{Decide}, \rulename{Sat}, \rulename{Consume$_1$}, \rulename{Consume$_2$}, \rulename{Unsat}, \rulename{T-Decide}, \rulename{T-Consume}, \rulename{Restart}), move clauses between any of the three places (\rulename{Propagate}, \rulename{Conflict}, \rulename{Backjump}, \rulename{Learn}) or remove existing clauses from the current state (\rulename{Forget}, \rulename{T-Backjump-Decide}).

\paragraph{\rulename{Resolve}.}

The \rulename{Resolve} rule takes the two clauses $C$ (the current conflict clause) and $D$ (from the trail $M$) and produces the resolvent with respect to some literal $L$.
By construction we know that \ResT has both $C$ and $D$ available and can thus apply the \rulename{Resolution} rule to produce the same resolvent.

\paragraph{\rulename{T-Propagate} and \rulename{T-Conflict}.}

Both the \rulename{T-Propagate} rule and the \rulename{T-Conflict} rule construct a new clause $E$ by calling the MCSAT $\Explain$ method.
This method produces ``a valid theory lemma'' (as specified in \cite{deMoura2013}), thus we have $T \models E$ and can obtain these clauses using the \rulename{Strong Theory Derivation} rule.

As we have simulated all MCSAT rules that can be used to construct new clauses, we can now take any MCSAT proof and convert it into a \ResT proof by using the presented simulations for the \rulename{Resolve}, \rulename{T-Propagate} and \rulename{T-Conflict} rule and removing all other proof steps.

We have shown that MCSAT and \ResT simulate each other and are thus equivalent (in terms of proof complexity), concluding our proof for \cref{thm:the-theorem}.

\section{Complexity of theory computations}
\label{sec:theory-computations}

One can very well argue that the length of a proof (in terms of proof steps) is not a satisfactory measure to assess the computational complexity of performing this proof.
We already see in the analysis above that we can make proof rules arbitrarily powerful so that we can essentially prove anything with a single step.
We could as well devise a prove system that executes individual processor instructions and thus requires an insane amount of proof steps.
To compare proof systems it is thus crucial that they operate on a similar level of abstraction as we assume every proof rule to take some fixed amount of time.

One such abstraction is that we assume that theory queries (whatever a \emph{theory query} may be in detail) only require constant effort here: one proof step for \rulename{Strong Theory Derivation} in \ResT and one proof step for either \rulename{T-Propagate} or \rulename{T-Conflict} in MCSAT.
One may very well be worried here, as we know that theory queries for some theories can be extremely expensive (even doubly exponential for non-linear arithmetic using cylindrical algebraic decomposition) while all the other rules are easy to compute.
This leads to two issues that influence how meaningful the above analysis is: how often are these particular rules used and do these theory queries have comparable computational effort?

If we use theory queries (almost) equally often in both proof systems, we can relax our assumption that all proof rules take constant time.
Instead we consider \emph{theory rules} (\rulename{Strong Theory Derivation}, \rulename{T-Propagate} and \rulename{T-Conflict}) and \emph{non-theory rules}.
We claim that all non-theory rules can be applied \emph{quickly} (low polynomial runtime) and it suffices to consider their overall number as we did above.
If we can additionally show that the theory rules have comparable computational effort in the different proof systems, this implies that the overall effort is (complexity-wise) the same.

Recalling the above reductions, we see that both proof systems need the exact same amount of theory rules.
MCSAT needs no theory rules to simulate \rulename{Resolution} and a single application of \rulename{T-Conflict} to simulate \rulename{Strong Theory Derivation}.
\ResT on the other hand only applies \rulename{Strong Theory Derivation} once to simulate \rulename{T-Conflict} or \rulename{T-Propagate}.

What remains to analyze is the computational effort for each of these theory queries.
When \ResT simulates MCSAT (as described above) it generates the same clauses that MCSAT generated: if we would have an implementation for \rulename{Strong Theory Derivation} that would do this (significantly) more efficiently than whatever MCSAT does in the $\Explain$ and $\Infeasible$ methods, we could take this method and use it in MCSAT as well.
We have also seen that MCSAT can simulate the \rulename{Strong Theory Derivation} rule (with a single theory call to the $\Infeasible$ method while $\Explain$ is trivial as no theory assignments are present \cite{deMoura2013}) and thus a more efficient $\Infeasible$ method would directly allow us to improve upon the implementation of the \rulename{Strong Theory Derivation} rule.
We can just assume that both proof systems may use the exact same techniques for the theory queries and thus their computational effort is the same.

\section{Impact on CDCL(T)}

From the perspective of the SMT community the comparison to \ResT is of course interesting, but the actual question is how MCSAT relates to CDCL(T).
Though we did not answer this specifically, we can both infer and speculate now more substantively than before.
We know from \cite{Robere2018} that CDCL(T) (with strong theory derivation) is equivalent to \ResT in terms of proof complexity as well, hence transitively MCSAT and CDCL(T) are equivalent (in terms of proof complexity).

CDCL(T) is presented as an algorithm in \cite{Robere2018}, but \cite{Nieuwenhuis2006} also (equivalently) defines CDCL(T) as a proof system.
Though we cannot claim \emph{equivalency} (beyond the aspect of proof complexity) between CDCL(T) and \ResT, we can very well do so between MCSAT and \ResT as argued above.
We conjecture that most proof rules from CDCL(T) and MCSAT nicely align and the ideas from the reductions above should be a good start for the remaining proof rules.
Thus we believe that this work might be a first step towards a proof that establishes equivalency between CDCL(T) and MCSAT.

Like for MCSAT the theory queries within CDCL(T) are much more specific than those from \ResT and thus we expect the issues discussed in \cref{sec:theory-computations} to essentially vanish here as well. Though the theory queries pose different questions, we know from practical implementations that the algorithms to answer them are very similar.

\section{Conclusion}

We have recalled the definitions of MCSAT (from \cite{deMoura2013}) and \ResT (from \cite{Robere2018}) and shown that they are equivalent with respect to their proof complexity, and even equivalent in a stronger sense that we described as \emph{algorithmically equivalent}.
We have discussed that though the complexity of theory queries is unknown, the fact that we simulate every theory query individually allows us to assume that the required effort is the same in both proof systems: if one would be (significantly) faster, we assume that the presented simulation can be used in the other proof system as well with only polynomial overhead.
Still a more thorough analysis would be important to conduct in the future.

Finally we have discussed the importance of the presented work for the actual goal, the comparison of MCSAT and CDCL(T).
We have noted that we established equivalency (in terms of proof complexity) of MCSAT and CDCL(T) which gives rise to the hope that they are in fact \emph{equivalent} in the sense that they can (almost) directly simulate each other using the ideas presented here.
A corresponding proof is also left for future work.

\bibliographystyle{alpha}
\bibliography{literature.bib}

\begin{thebibliography}{BHvMW09}

\bibitem[BHvMW09]{handbook}
A.~Biere, M.~Heule, H.~van Maaren, and T.~Walsh.
\newblock {\em Handbook of Satisfiability}, volume 185 of {\em Frontiers in
  Artificial Intelligence and Applications}.
\newblock IOS Press, 2009.

\bibitem[DLL62]{Davis1962}
Martin Davis, George Logemann, and Donald Loveland.
\newblock A machine program for theorem-proving.
\newblock {\em Communications of the ACM}, 5:394--397, 1962.

\bibitem[dMJ]{deMoura2013}
Leonardo de~Moura and Dejan Jovanović.
\newblock A model-constructing satisfiability calculus.
\newblock In {\em Proceedings of VMCAI 2013}, volume 7737 of {\em LNCS}, pages
  1--12.

\bibitem[JdM]{Jovanovic2012}
Dejan Jovanovi{\'c} and Leonardo de~Moura.
\newblock Solving non-linear arithmetic.
\newblock In {\em Proceedings of IJCAR 2012}, pages 339--354.

\bibitem[KS08]{decision_proc}
Daniel Kroening and Ofer Strichman.
\newblock {\em Decision Procedures: {A}n Algorithmic Point of View}.
\newblock Springer, 2008.

\bibitem[MS]{Silva1996}
{Jo\~{a}o}~P. {Marques Silva} and Karem~A. Sakallah.
\newblock Grasp: a new search algorithm for satisfiability.
\newblock In {\em Proceedings of ICCAD 1996}, pages 220--227.

\bibitem[NOT06]{Nieuwenhuis2006}
Robert Nieuwenhuis, Albert Oliveras, and Cesare Tinelli.
\newblock Solving {SAT} and {SAT} modulo theories: From an abstract
  {D}avis--{P}utnam--{L}ogemann--{L}oveland procedure to {DPLL}(\emph{T}).
\newblock {\em Journal of the {ACM}}, 53:937--977, 2006.

\bibitem[RKG]{Robere2018}
Robert Robere, Antonina Kolokolova, and Vijay Ganesh.
\newblock The proof complexity of {SMT} solvers.
\newblock In {\em Proceedings of CAV 2018}, volume 10982 of {\em LNCS}, pages
  275--293.

\end{thebibliography}

\end{document}